# Magnetic proximity effect on excitonic spin states in Mn-doped layered hybrid perovskites

*Timo Neumann[1,2], Sascha Feldmann[1], Philipp Moser[2], Jonathan Zerhoch[2], Tim van de Goor[1], Alex Delhomme[3], Thomas Winkler[1], Jonathan J. Finley[2], Clément Faugeras[3], Martin S. Brandt[2], Andreas V. Stier[2], Felix Deschler[2,\*]*

[1]Cavendish Laboratory, University of Cambridge, Cambridge, UK

[2]Walter Schottky Institut and Physik Department, Technische Universität München, Garching, Germany

[3]Université Grenoble Alpes, INSA Toulouse, Univ. Toulouse Paul Sabatier, EMFL, CNRS, LNCMI, Grenoble, France

**Abstract**

Materials combining the optoelectronic functionalities of semiconductors with control of the spin degree of freedom are highly sought after for the advancement of quantum technology devices. Here, we report the paramagnetic Ruddlesden-Popper hybrid perovskite Mn:(PEA)$_2$PbI$_4$ (PEA = phenethylammonium) in which the interaction of isolated Mn$^{2+}$ ions with magnetically brightened excitons leads to circularly polarized photoluminescence. Using a combination of superconducting quantum interference device (SQUID) magnetometry and magneto-optical experiments, we find that the Brillouin-shaped polarization curve of the photoluminescence follows the magnetization of the material. This indicates coupling between localized manganese magnetic moments and exciton spins via a magnetic proximity effect. The saturation polarization of 15% at 4 K and 6 T indicates a highly imbalanced spin population and demonstrates that manganese doping enables efficient control of excitonic spin states in Ruddlesden-Popper perovskites. Our finding constitutes the first example of polarization control in magnetically doped hybrid perovskites and will stimulate research on this highly tuneable material platform that promises tailored interactions between magnetic moments and electronic states.

**Main text**

The development of materials which are simultaneously magnetic and semiconducting, while retaining excellent opto-electronic properties and high luminescence yields, is a scientific challenge, which holds great potential for creating novel opto-spintronic functionality for information- and communication technologies.[1–3] Dilute magnetic semiconductors (DMS) constitute a material class which combines these properties by introducing magnetic impurities to an otherwise non-magnetic host semiconductor.[4] Inorganic DMS, most prominently transition metal-doped III-V semiconductors, have been known for decades and have enabled control over various processes, such as highly circularly polarized photoluminescence (PL),[5] spin injection,[6] giant magnetoresistance[7] and control of magnetism by electric fields and currents.[8,9] However, the demanding -mostly epitaxial- material processing techniques as well as limiting material tunability diminish applicability beyond fundamental research.[10–12] Metal-halide perovskites offer an opportunity for control of spin in a high-performance semiconductor due to their exceptional tolerance to structural defects and impurities,[13,14] combined with their ease of production as polycrystalline thin films and nanostructures via simple scalable solution-processing techniques.[15] They have created disruptive changes in the field of solution-processed semiconductors for optoelectronics, where they are used in highly efficient energy conversion applications, such as solar cells and LEDs.[16–18] High luminescence yields are an exceptional property of the perovskites,[19] which are preserved under changes in chemical composition and structure, for example by halide-mixing for tuning of the optical band gap over the visible range.[20] We now show that they hold promise for exciting opportunities in the field of opto-spintronics.[21,22] So far, transition metal-doping has proven useful for altering the optoelectronic properties of lead halide perovskites, e.g. by causing enhanced luminescence in Mn-[23], Ni-[24] and Cu-[25] doped nanocrystals, and modified growth and solar cell performance in Mn-, Fe-, Co-, Ni-doped thin films.[26,27] Single crystals of MA(Mn:Pb)$I_3$ (MA = methylammonium) show ferromagnetism ($T_C$ = 25 K) which is optically switchable,[28] indicating coupling between the dopants' *d*-electron spins and optically excited charge carriers.

In this work we report the paramagnetic manganese-doped 2D Ruddlesden-Popper perovskite Mn:(PEA)$_2$Pb$I_4$ (PEA = phenethyl ammonium, from now on Mn:PEPI and PEPI for the doped and undoped material, respectively) and study its luminescent properties under magnetic fields. The PL from the magnetically brightened emissive state shows circular polarization up to 15%, which we find to be directly proportional to the material's magnetization. We attribute this effect to a spin-orienting dipole-dipole interaction caused by proximity of manganese spins and localized excitons. These findings constitute the first demonstration of exciton spin control in a transition metal-doped lead-halide perovskite and provide a first step towards opto-spintronic functionalities of these materials.

*Fabrication and Characterisation of Magnetic Ruddlesden-Popper Hybrid Perovskites*

We fabricate Mn:PEPI hybrid perovskite films (Fig. 1a) using an established protocol for PEPI[29] with additional Mn precursor (1% atomic ratio Mn relative to Pb). X-ray diffraction measurements and fit to the PEPI crystal structure[30] reveal that Mn-doped and undoped samples are of high crystallinity, showing the typical diffraction pattern of a highly oriented (PEA)$_2$PbI$_4$ film, where the layers of the crystal domains are aligned parallel to the substrate (Fig. 1b). Absorption and PL spectra show a sharp excitonic resonance at 2.394 eV and a narrow emission peak at 2.353 eV, respectively (Fig. S1a). The Mn-doped sample absorption and emission are blue-shifted by 10 meV and 5 meV, respectively. This is caused by a small bromide content, since MnBr$_2$ was used as the Mn-precursor for solubility reasons. To investigate the magnetic properties of the samples, we carried out electron paramagnetic resonance (EPR) measurements at 4 K (Fig. 1c). We detect a weak response for the pristine sample, likely due to very small amounts (ppm) of paramagnetic impurities, showing that the fabricated undoped hybrid perovskites are diamagnetic (Fig. S2a). Upon introducing Mn to the material, a strong and broad resonance appears around a field of 330 mT, corresponding to a g-factor of ~2. On top of the single broad resonance, a clear Mn sextet is observed, which occurs due to the hyperfine coupling of the manganese's $S = 5/2$ electron spin to its nuclear spin of $I = 5/2$. The distinct character of the hyperfine resonances indicates the existence of magnetically isolated Mn$^{2+}$ in the sample, since strong broadening due to interactions occurs in the case of Mn ions in close proximity.[31–33] To confirm the magnetic nature of Mn:PEPI perovskite, the magnetization of the sample was evaluated with varying temperature and magnetic field using SQUID-magnetometry. For the applied fields of 0.01 T, 0.1 T and 1 T the sample shows a 1/T Curie-law dependence on the temperature with no difference between heating and cooling, as is expected for paramagnetic materials (Fig. S2c).[34] The magnetization with magnetic field sweep does not show hysteresis and follows the Brillouin function for non-interacting $J=5/2$ spin systems, corresponding to the spin alignment of the high-spin Mn $d^5$ configuration in the magnetic field (Fig. 1d). We conclude that we successfully induced paramagnetism to the diamagnetic PEPI perovskite by adding small amounts of MnBr$_2$ to the precursor solution. In the following, we investigate the effect of local magnetization on excitonic state recombination in this novel material.

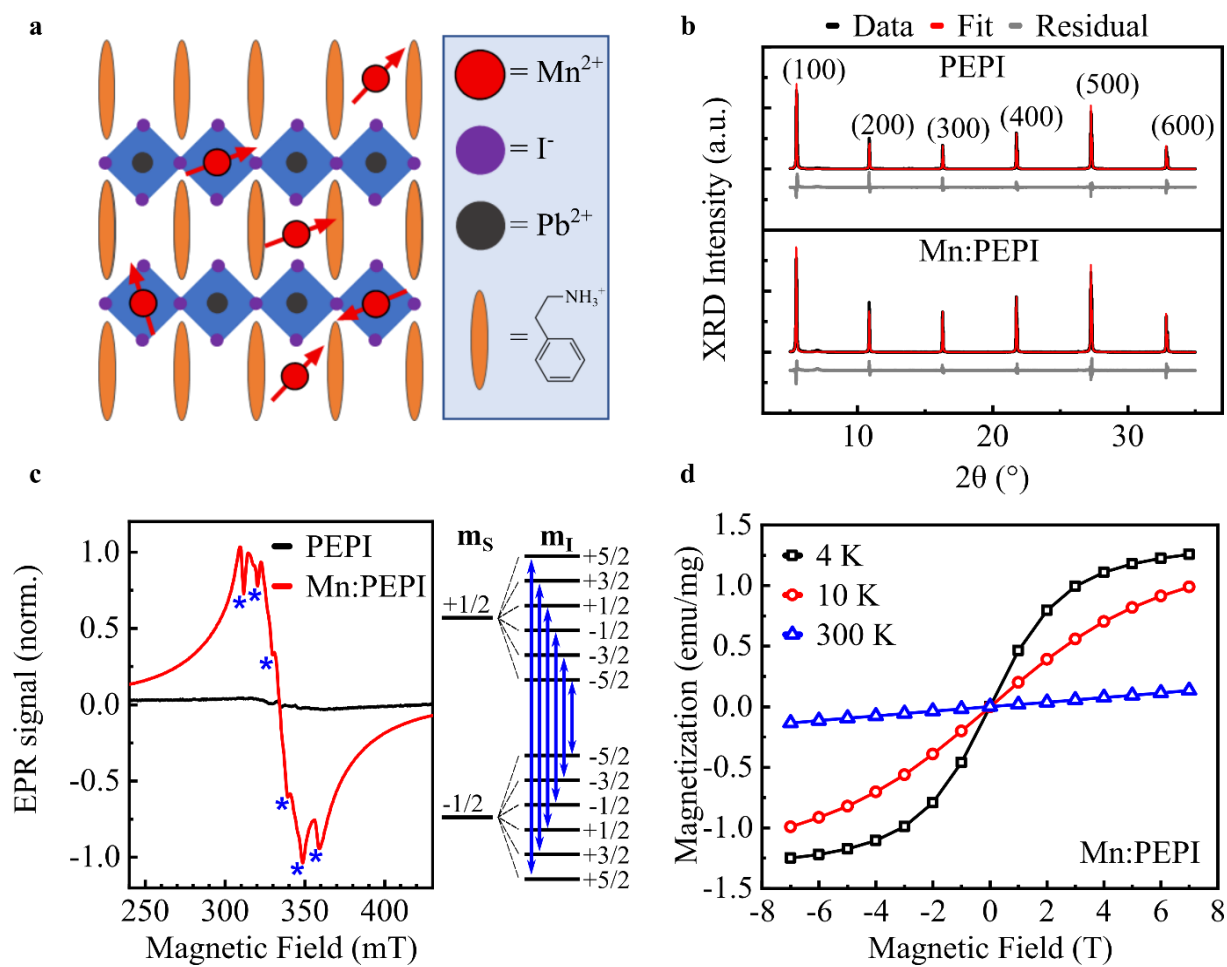

**Figure 1: Structural and magnetic characterization of Mn:PEPI. a**, Sketch of the Ruddlesden-Popper quantum well structure with two possible Mn sites: substitutional for the divalent anion and interstitial in the organic barrier. **b**, X-ray diffraction pattern of perovskite films. **c**, Electron paramagnetic resonance response of drop casted material. Stars denote the six hyperfine transitions. **d**, SQUID magnetization versus field sweep at different temperatures. The data shows that Mn$^{2+}$ ions are present in the crystal structure and give rise to paramagnetism.

*Magnetic brightening of dark localized exciton*

Temperature-dependent PL spectroscopy was employed to investigate the emission spectrum of the materials at low temperatures. Upon cooling from room temperature, the excitonic peak undergoes a redshift, while two additional peaks emerge below 100 K and 10 K (Fig. S1b). At 4 K, the emission spectrum shows three features at 2.342 eV, 2.334 eV and 2.302 eV, respectively. While the high-energy peak has been well-established as the free exciton emission, the lowest energy peak has been proposed originate from self-trapped excitons, phonon replicas, exciton-polarons or an out-of-plane oriented magnetic dipole transition.[35–38] Since this peak is of no importance for the further findings of this report, we will not discuss it in more detail. The middle peak at 2.334 eV, which is barely visible without magnetic field, has not been discussed in the literature. Studies on $(PEA)_2PbBr_4$, where a similar spectrum was observed, concluded that this peak corresponds to a dark localized exciton which gains oscillator strength by mixing with a bright state via spin-orbit coupling.[39] In the following, we will refer to the two relevant emission peaks at 2.342 eV and 2.334 eV as the (bright) free exciton and (dark) localized exciton, respectively.

When a magnetic field is applied orthogonal to the quantum well plane, PL spectra from pristine and doped perovskite exhibit strong field-dependent effects in intensity and polarization, with different field-response of the two peaks (Fig. 2). The free exciton peak intensity decreases with magnetic field and a clear asymmetry between the direction of the field is observed (Fig. 2a, 2b, Fig. 3a top), indicating a circularly polarized emission with a near linear dependence (~1.2% / T, Fig. S4b) on applied magnetic field. This magnetic field effect with a similar slope has been reported in $MAPbI_3$, where it was attributed to magnetic-field-induced spin-mixing of the photogenerated electron–hole pairs with different g-factors.[40] Furthermore, two new peaks appear approximately 1 meV blue- and red-shifted from the zero-field peak, possibly due to a splitting with magnetic fields.[40] However, due to the broad linewidth of the strongly overlapping peaks, this effect cannot be analysed in greater detail within this study.

Notably, the localized exciton emission greatly increases with magnetic field to an approximately four-fold enhanced intensity at 14 T, compared to zero-field. For the undoped sample, no difference in intensity or peak position with field direction is observed and the circular polarization remains zero at all fields (Fig. 2a, 2b, Fig. 3a bottom). Magnetic brightening of dark states is a known effect in magneto-spectroscopy.[41–43] Commonly it originates from two different mechanisms which are either the shrinking of the excitonic wave function and enhancement of the oscillator strength, or the mixing of spin states allowing the dark state to 'borrow' oscillator strength from a bright state via magnetic coupling.[44] In both scenarios the magnetic field directly affects the recombination mechanism and decreases the PL lifetime rather than affecting the population of states by shifting of their energies. In monolayer $WSe_2$ the PL intensity increases quadratically with magnetic field up to fields above 30 T, which is a hallmark for brightened emission from a spin-forbidden free exciton transition.[45,46] In contrast, an initially linear increase with saturation at higher fields is observed in PEPI (Figure 2b inset), possibly due to the filling of the finite number of localized states. While the PL intensity of the localized exciton peak is greatly enhanced with magnetic field, the intensity of the free exciton peak decreases. This suggests that the magnetic field increases either the bright to dark exciton transfer rate (i.e. increasing the population of the dark excitons) or the dark exciton radiative recombination rate. Resolving the detailed photophysical mechanism requires transient spectroscopies under applied magnetic field and will be investigated in future work. Similar observations were made in experiments on single $FAPbBr_3$ nanocrystals, where magnetic brightening was employed to brighten the dark singlet state and reveal the energetic order of exciton states as a dark singlet ground state below a bright triplet.[47] However, the overall emission intensity of those nanocrystals at low temperature is barely affected by the magnetic brightening of the dark singlet ground state, since the lack of an efficient relaxation mechanism leaves said state only sparsely populated. In contrast, the strong absolute increase of PL emission in PEPI shows that the dark state is densely populated and constitutes a major loss channel. This strongly limits the PL quantum efficiency at room temperature, potentially also in similar two-dimensional perovskite systems.

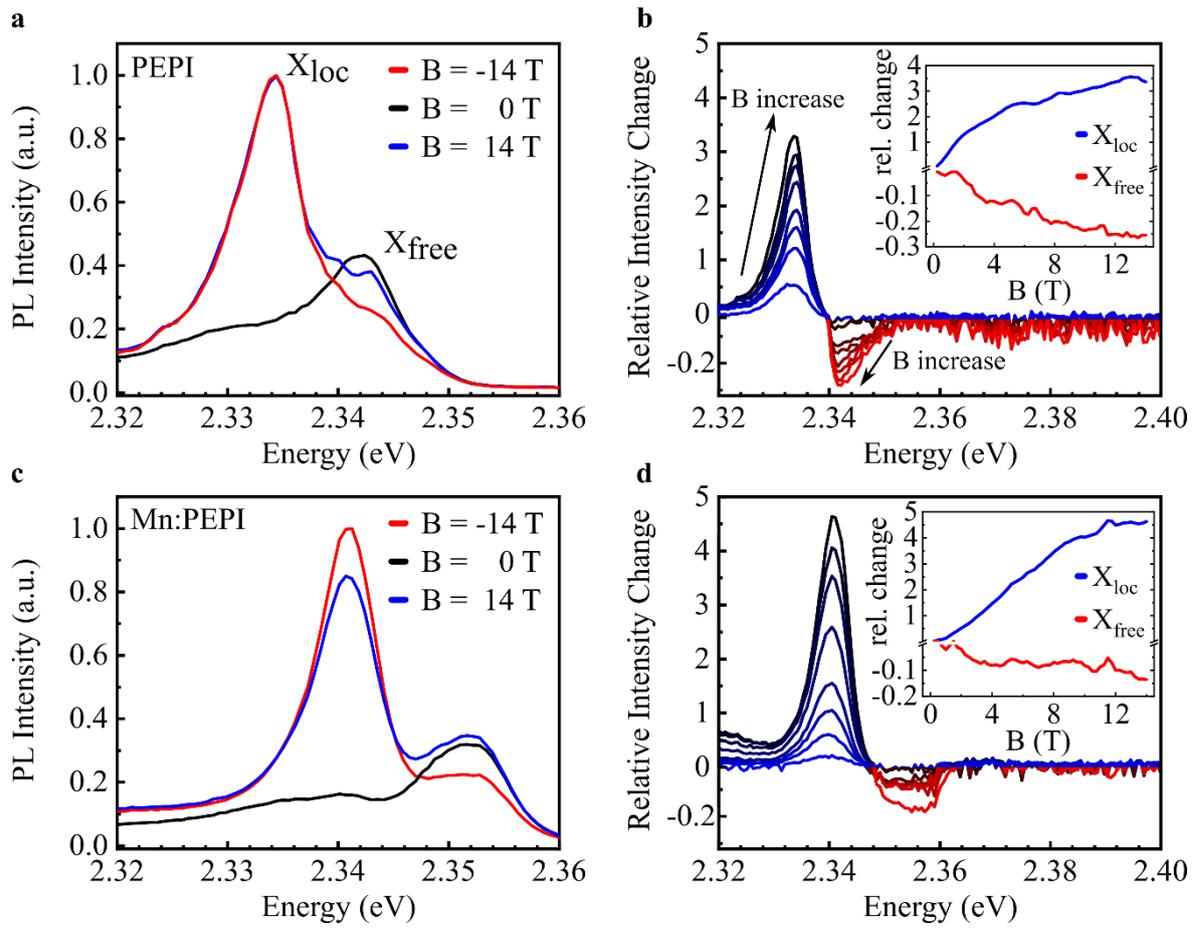

**Figure 2: Low-temperature magneto-photoluminescence spectra of PEPI and Mn:PEPI. a** and **c**, circularly polarized emission spectra at 4 K of the free and localized exciton with continuous wave excitation at 395 nm. Measurements were performed in Faraday geometry with B = 14 T corresponding to σ+ detection. **b** and **d,** relative unpolarized intensity change with increasing magnetic field from 1 T to 14 T, for PEPI and Mn:PEPI, respectively. Insets: relative intensity change of the free and localized exciton peak.

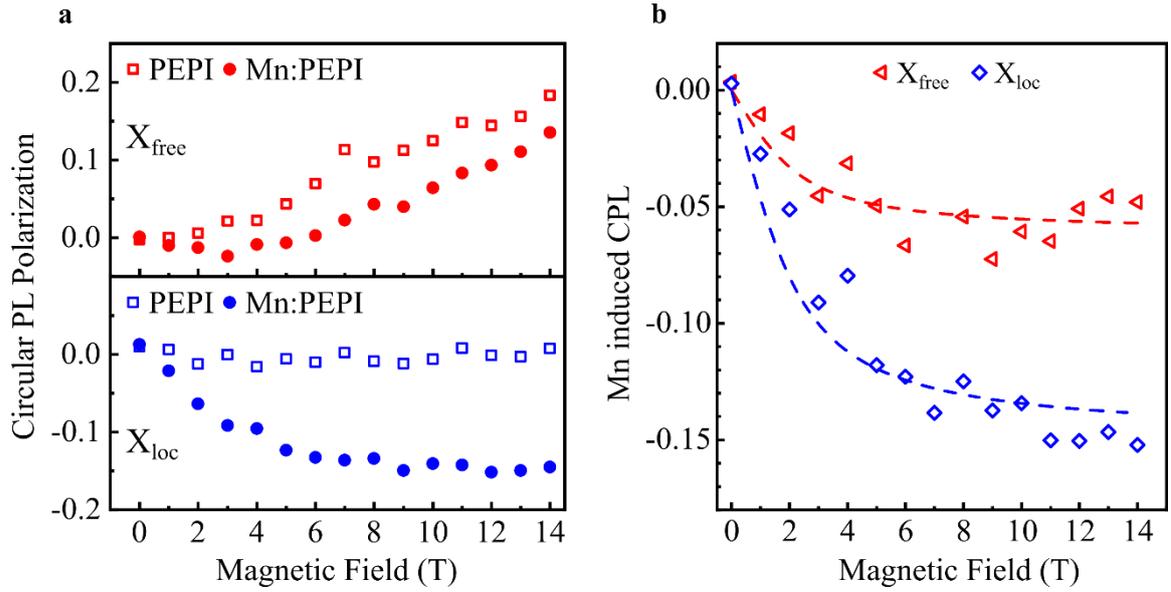

**Figure 3: Circular PL polarization of PEPI and Mn:PEPI. a**, Comparison of undoped PEPI and doped Mn:PEPI for the free (top) and localized (bottom) exciton emission. **b**, Difference of circular polarization of undoped and doped perovskite for the free and localized exciton. Dashed lines are Brillouin function ($T = 4$ K, $J = 5/2$) fits yielding a saturation polarization of ~6% and ~15% for the free and localized exciton, respectively.

*Exciton polarization control via manganese doping*

The magneto-PL of the doped Mn:PEPI perovskite (Fig. 2c, 2d) shows a similar magnetic brightening of the localized exciton as the undoped sample. However, we find that the polarization of the PL is greatly affected by the presence of the magnetic dopants. A clear difference in intensity between the positive and negative magnetic field directions is observed, corresponding to a circular polarization (CP) of the PL (Fig. 2c). The CP increases nearly linearly for fields up to 2 T, then it starts to saturate and reaches a constant value (Fig. 3a bottom). Comparison of the absolute slope in the linear region with that of the free exciton in the undoped material (~3.0% / T vs. ~1.2% / T, Fig. S4b) yields a two-and-a-half-fold increase and shows that manganese-induced circular polarization can be tuned more efficiently than the intrinsic field-induced polarization. The polarization of the free exciton shows qualitatively the same behaviour, but weaker in magnitude and on top of the linear trend that also occurs in the nonmagnetic control sample. In order to solely show the additional polarization induced by the magnetic doping, the difference between PEPI and Mn:PEPI is depicted in Fig. 3b. Both free and localized exciton emission show the same polarization behaviour, described by a Brillouin function for $J = 5/2$ spins at 4 K with saturation polarization of 6% and 15%, respectively. The observed effect is independent of the excitation polarization and antisymmetric with respect to the magnetic field direction (Fig. S3). Superimposing both polarization curves with the magnetization of the sample and the Brillouin function reveals a direct proportionality between the alignment of the magnetic $Mn^{2+}$ dopants' magnetic moment and the polarization of the PL of the paramagnetic perovskite (Fig. 4a). This proportionality unambiguously shows the coupling of manganese spins and the host semiconductor's excitonic states and confirms that magnetic doping provides a viable approach to control exciton spin states in perovskites.

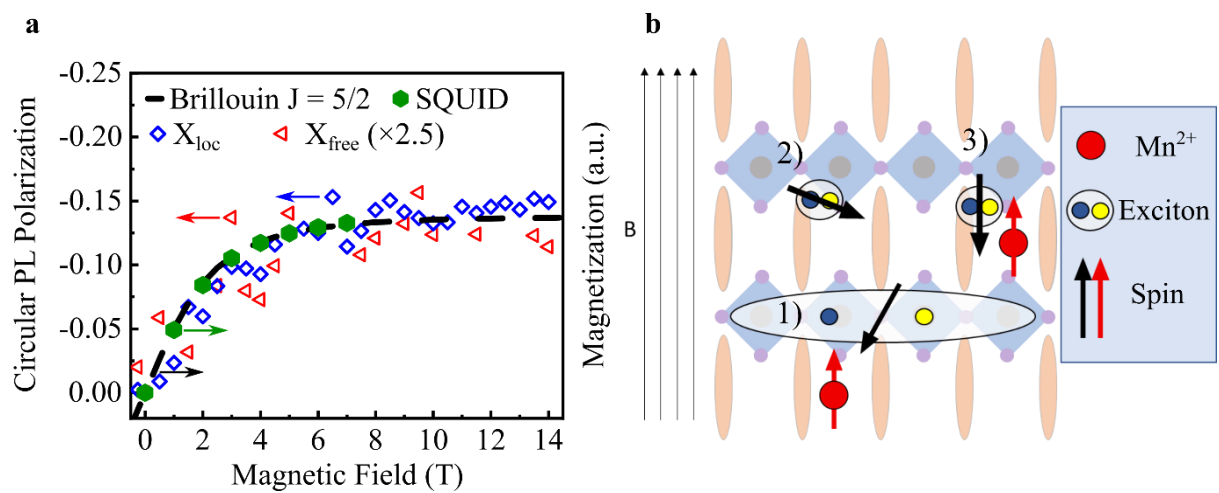

**Figure 4: Direct relation between magnetization and PL polarization behaviour in Mn:PEPI and proposed interaction mechanism. a**, Superposition of PL polarization and magnetization curve of Mn:PEPI at 4 K and Brillouin function for a non-interacting $J = 5/2$ spin system. **b**, Schematic of the proposed interaction between excitons and aligned manganese spins in an external magnetic field. Although exact manganese and exciton location are yet to be confirmed, we depict an interstitial Mn and interfacial exciton site to illustrate the potential mechanism. Three possible cases are distinguished: 1) free exciton, 2) exciton localized far from manganese dopant, 3) exciton localized close to manganese dopant.

*Mechanism for Mn-doping induced PL polarization in PEPI*

While magnetic transition metal-doping is an established technique to induce magnetic properties in inorganic semiconductors, the observed emission behaviour of the magnetic perovskite only partially reflects the characteristics of classical dilute magnetic semiconductor quantum wells. Arguably the most significant feature of a magnetic semiconductor is a giant Zeeman effect where the effective g-factor of the charge carriers can be enhanced by as much as two orders of magnitude due to the direct exchange between d-electrons of the localized dopant and the conduction/valence band of the host semiconductor (sp-d exchange).[12,48,49] The Zeeman energy shift between the different spin-sub-bands is proportional to the magnetization of the material and often several tens of meV large.[50,51] With such large energy differences, a large and often near unity PL polarization arises from the fast relaxation to the lowest-energy spin-polarized state, and thus, a greatly imbalanced thermal population of spin states. Although the polarization of the Mn:PEPI follows the magnetization of the material, no (giant) Zeeman splitting of the PL peaks is observed for the doped (Fig. S4a) and undoped sample. We conclude that the sp-d exchange is negligible in Mn:PEPI and does not explain the induced polarization. This observation is in contrast to several transition metal-doped nanostructures, which show polarization of either the excitonic or the Mn d-d transition as well as giant Zeeman splitting caused by the sp-d exchange.[5,52,53]

Instead, we propose the alignment of exciton spins with respect to the manganese spins as explanation for the presented results, due to dipole-dipole interaction if both species are in close proximity. Studies on nonmagnetic CdSe quantum dots separated from a magnetic semiconductor quantum well by a spacer layer compared the emission characteristics of both coupling mechanisms within the same heterostructure.[53] While the emission peak from the magnetic semiconductor shows giant Zeeman splitting and unity polarization at low fields, the emission from the quantum dots show no Zeeman splitting and a weaker polarization, depending on the distance between magnetic and nonmagnetic semiconductor.

This spin-aligning magnetic proximity effect provides a plausible mechanism to explain the much greater effect of manganese doping on the polarization of the localized exciton than on the free exciton, if localization near the Mn site is assumed (Fig. 4b). In this case, the imbalance of exciton spin population and accordingly of the PL polarization is given by the fraction of excitons that are localized near a Mn binding site as well as the average interaction distance. Since the free exciton is delocalized over several unit cells and confined to the $[PbI_6]^{4-}$ layer, the distance to the Mn ion is on average large and the interaction weak. Here, exciton spins are mostly randomly oriented with the weak coupling to the Mn causing only a small polarization, proportional to the alignment of the Mn in the applied magnetic field. In contrast, a localized exciton is much closer to the Mn ion enabling strong interactions and highly polarized emission from said state. Maximizing this effect requires precise knowledge of the Mn site and the exciton localization site within the perovskite lattice, both of which are still unknown and topic of future investigations.

The substitution of Pb by Mn has been shown for bulk $CsPbBr_3$ and $CsPbCl_3$ by a combined EPR and NMR (nuclear magnetic resonance) study,[54] but applicability to hybrid Ruddlesden-Popper systems is not evidently given. In other reports about Mn doped perovskite nanocrystals[23] or bulk $MAPbI_3$[28] and $(BA)_2PbBr_4$ (BA = butyl ammonium),[55–57] substitution was also proposed, but interstitial lattice sites, surface decoration or phase segregation, as observed for other dopants, cannot be excluded.[54,58,59] On the other hand, the flexibility of the organic layer between two quantum wells has been shown to allow for intercalation of large molecules like halogens and chlorobenzene,[60–62] as well as electrochemical intercalation of $Li^+$ ions.[63] Although the specifics of these materials and experiments make their results not directly transferable, they imply the possibility of Mn intercalation in the organic layer as a peculiarity of Ruddlesden-Popper perovskites. Regarding the exciton localization site, spectroscopic studies on 4-methylbenzylammonium lead bromide observed emission from a localized exciton,[64] potentially linked to differences in the local structure at the inorganic-organic interface, which were investigated by NMR studies on $(PEA)_2PbBr_4$ and $(PEA)_2PbI_4$.[65,66] More recent studies have also reported the localization of excitons at the interface,[67] a phenomenon well known in inorganic semiconductor quantum well structures.[68,69] We note that the combination of a localized interfacial exciton and an interlayer Mn binding site are supportive of our proposed interaction mechanism and can explain the differences in coupling strength between the Mn dopant and the two discussed excitonic states. In a comparative study on manganese-doped $CsPb(Cl,Br)_3$ nanocrystals, which exhibit only one excitonic emission peak (Fig. S5), we observe Zeeman splitting as well as circularly polarized emission as a linear function of magnetic field. However, the magnetic field effects in these nanocrystals are not affected by manganese doping, stressing the importance of the nature of excitonic states in the 2D layered structure. While the proposed model qualitatively describes the observed magneto-PL spectra, it will require further testing, for example with near edge x-ray absorption spectroscopy to confirm the dopant location within the lattice. We hope to stimulate research about other important factors, such as the effect of exciton triplet/singlet nature on the magnetic coupling and the confinement[70] and structure of the exciton wavefunction,[71] which must be considered for a comprehensive model.

In summary, to our best knowledge, the here reported Mn:PEPI constitutes the first lead-halide perovskite material in which magnetic doping enables direct control of exciton spin states via a magnetic field. In contrast to epitaxially fabricated magnetically doped semiconductors, magnetic dopants are easily introduced during fabrication from solution processing. Considering the structural and energetical tunability of the perovskite materials, we envision future opportunities for magnetic semiconductors with tailored exciton-dopant interaction and a range of potential applications in spin-based technologies.

**Materials & Methods**

*Sample fabrication:* All steps of the sample fabrication were carried out in a nitrogen filled glovebox. The Pb (Mn) precursor solutions were prepared by dissolving $PbI_2$ (LumTech, >99.999%) or $MnBr_2$ (Sigma, >98%) and phenethyl ammonium iodide (LumTech, >99.5%) salts in N,N-dimethylformamide (Sigma, >99.8% anhydrous) with a molar ratio of 1:2. The solutions were stirred for 2 h at 80°C and filtered with a 0.2 µm pore size PTFE syringe filter. The Pb and Mn precursors were mixed in an atomic ratio of 99:1 to yield the Mn:PEPI solution. Films were prepared on oxygen-plasma-treated glass coverslips by spin coating at 2000 rpm for 30 s and by drop casting and solvent evaporation at 120°C for 2 h. Samples for the optical measurements were encapsulated.

*EPR:* Electron paramagnetic resonance measurements were carried out on drop casted samples. The setup consisted of a Jeol RE series ESR spectrometer with Jeol JES RE2X power supply and field controller, a Jeol X-Band microwave source and a $TE_{102}$ (Bruker ER 4102ST) resonator with 100 kHz modulation unit. The temperature in the Oxford ESR 900 cryostat was controlled by a Lake Shore 335 temperature controller. A Stanford Research System Model SR830 DSP lock-in amplifier was used.

*SQUID:* Magnetic properties were measured on a Quantum Design Magnetic Properties Measurement System MPMS XL-7. The sample contained 20 mg of drop casted material in a medical capsule. The sample was cooled from 300 K to 2 K with a rate of 10 K/min and the magnetic field was swept -7 T to 7 T with a rate of 0.2 T/min. Before each measurement point, the system was given one minute to equilibrate.

*Magneto-PL:* Magneto photoluminescence measurements were carried out in the LNCMI-G (Laboratoire National des Champs Magnétiques Intenses - Grenoble). The samples were installed in a closed tube filled with helium exchange gas and cooled down to 4 K. The excitation source was a 395 nm (3.139 eV) diode laser guided through a fiber and focused onto a ~5 µm spot with an excitation power of 1 µW (5.09 W/cm$^{-2}$) on the sample. In the detection path, a right-handed polariser ($\sigma^+$) was mounted. The magnetic field was swept from 14 T to -14 T with a step size of 0.25 T. For each magnetic field, three spectra were obtained with an integration time of 10 s per spectrum.

*XRD:* X-ray diffraction measurements were made on thin film samples on a Bruker D8 discover diffractometer with Cu Kα radiation, λ = 1.5403 Å. Samples were measured using a Bragg-Brentano geometry over $5 \leq 2\theta \leq 65$ with a step size of $\Delta 2\theta = 0.001$. Patterns were fitted using the LeBail method in TOPASv5. The background was modeled with a Chebyshev polynomial function of 3rd order and the peak shape was set to Thompson-Cox-Hasting pseudo-Voigt. In order to account for the strong preferential orientation, only the c lattice parameter and zero offset were refined.

*UV/VIS and PL:* Absorption spectra of spin coated films were measured using a Shimadzu UV600 spectrometer in linear transmission mode. PL spectra were measured on an Edinburgh Instruments FLS90 fluorimeter.

**Author contributions**

T.N., S.F. and F.D. conceived and planned the experiments. T.N. fabricated the samples, performed X-ray diffraction, SQUID magnetometry, absorption spectroscopy and initial magneto-spectroscopy with input from S.F., T.v.d.G. and T.W. T.v.d.G. fitted the XRD data. J.Z performed EPR measurements with input from M.S.B. P.M., A.D, A.S. and C.F. performed low temperature magneto-PL spectroscopy at LNCMI Grenoble. T.N. and F.D. drafted the manuscript and compiled figures, with discussion of results and feedback on the manuscript from all authors.

**Acknowledgments**

T.N. acknowledges funding from the Winton Programm for the Physics of Sustainability. S.F. acknowledges funding from the Studienstiftung des deutschen Volkes and EPSRC, as well as support from the Winton Programme for the Physics of Sustainability. A.V.S. and P.M. gratefully acknowledge the German Science Foundation (DFG) for financial support via the Cluster of Excellence e-conversion EXS 2089. T.v.d.G. acknowledges support from the EPSRC Cambridge NanoDTC, EP/L015978/1. Part of this work was performed at the LNCMI, a member of the European Magnetic Field Laboratory (EMFL). F.D. acknowledges funding from the Winton Programm for the Physics of Sustainability, the DFG Emmy Noether Program and an ERC Starting Grant. This project has received funding from the European Research Council (ERC) under the European Union's Horizon 2020 research and innovation programme (grant agreement No 852084.